# Oxygen disorder in ice probed by X-ray Compton scattering.

**Ch. Bellin,[1] B. Barbiellini,[2] S. Klotz,[1] T. Buslaps,[3] G. Rousse,[1] Th. Straessle,[4] A. Shukla[1]**

[1]Institut de Minéralogie et Physique de la Matière Condensée, Univ. Paris VI, UMR CNRS 7590, case 115, 4 pl. Jussieu, 75252 Paris cedex 05, France

[2]Physics Department, Northeastern University, Boston, Massachusetts 02115, USA

[3]European Synchrotron Radiation Facility (ESRF), BP 220, 38043 Grenoble-Cedex, France

[4]Laboratory for Neutron Scattering, ETH Zurich and Paul Scherrer Institute, WHGA 144, CH-5232 Villigen PSI, Switzerland

**Abstract.**

We use electron momentum density in ice as a tool to quantify order-disorder transitions by comparing Compton profiles differences of ice VI, VII, VIII and XII with respect to ice Ih. Quantitative agreement is found between theory and experiment for ice VIII, which is the most ordered phase. Robust signatures of the oxygen disorder are identified in the momentum density for the VIII-VII ice phase transition. The unique aspect of this work is the determination of the fraction $n_e$ of electron directly involved in phase transitions as well as the use of position space signatures for quantifying oxygen site disorder.

# I. INTRODUCTION

Water exhibits a wide range of crystalline and amorphous ices. To date at least 15 crystalline phases and 3 amorphous phases [1] have been identified. The richness and complexity of the phase diagram of water stems from the hydrogen bond which is solely responsible for the formation of condensed phases [2] and the related structural versatility of the tetrahedral packing and proton disorder. The intermediate strength of the hydrogen bond permits a large number of possible variations for bonding in terms of angles and bond lengths, as observed for ices obtained at intermediate pressures below 1 GPa, i.e. ices II, III, IV, V, VI and IX. Among these, the greatest distortions in bond angles and lengths are observed for ice VI and can be traced to its peculiar structure. It is composed of two interpenetrating hydrogen-bonded networks, as is the case for ices produced at higher pressures such as ices VII and VIII. In other terms, ice VI carries features common both to ices made at intermediate and at higher pressure (it has a triple point with ices VII and VIII obtained above 1 GPa). Ice XII [3, 4] exists within the ice VI phase space and is composed of non-interpenetrating seven- and eight molecule rings [5]. The structures of both ice VII and ice VIII are composed of two interpenetrating tetrahedrally bonded networks.

In this paper, we report on Compton scattering measurements of some of the highest mass density ice structures, i.e. ices VI, VII, VIII and XII as well as ice Ih used only as reference. Apart from ice VIII, our studied ices are proton disordered while ice VII has an additional oxygen disorder which distinguishes it from otherwise identical ice VIII [6]. This site disorder is not yet fully understood [7, 8], and *ab initio* calculations have tried to shed some light on the problem [9]. Our measurements allow us to calculate the fraction of electrons directly involved in the different ice phase transitions. This is done by measuring accurately the number of 'displaced' electrons using the change in shape of a measured Compton profile across various phase transitions. It is made possible by the fact that the area of the profile, corresponding to the total number of electrons, remains constant. X-ray Compton scattering thus provides the opportunity to quantitatively study the phase diagram of ice. In particular we provide a new insight into the role of oxygen disordering in ice VII when compared to ice VIII.

## II. EXPERIMENTAL PROCEDURE AND CALCULATIONS

The Compton profile within the impulse approximation [10] is given by:

$$J(q, \mathbf{e}) = \int n(\mathbf{p})(\mathbf{p}.\mathbf{e} - q)d\mathbf{p} = \sum_{\substack{\text{occupied} \\ \text{states}}} \int \chi^*(\mathbf{p})\chi(\mathbf{p})\delta(\mathbf{p}.\mathbf{e} - q)d\mathbf{p} \quad (1)$$

where $\mathbf{e}$ is the unit vector along the scattering vector $\mathbf{K}$, $n(\mathbf{p})$, the electron momentum density and $\chi(\mathbf{p})$, the electron wave function in momentum space, i.e. the Fourier transform of the wave function in real space [11, 12]. Throughout the remainder of this paper we shall use atomic units (a. u.), for which $\hbar=m=1$. X-ray Compton scattering has been shown to be of particular importance in studying bonding in water and ice [13, 14, 15, 16, 17]. The technique is complementary to diffraction techniques, which are principally sensitive to ionic positions. Compton scattering is particularly sensitive to valence electron wave functions and to chemical bonding and as a consequence well suited to the study of hydrogen bonding. The information provided by Compton scattering about the ground-state electron distribution can be directly related to the Fourier transform of the real-space wave functions, so that one can track the changes in the electronic distribution under a variety of thermodynamic conditions, relate them to the changes in bonding and ultimately to the amount of hydrogen bonds in water [17, 18, 19, 20, 21].

The quenched samples were prepared using a Paris-Edinburg pressure cell. For each ice sample, approximately 100 mm$^3$ of distilled de-ionized water was loaded in the cell. Ice VI was produced by compressing $H_2O$ to 1.5 GPa. By cooling ice VI and compressing it up to 5 GPa below 95 K, pure ice VII was produced as described in ref. [6]. Ice VIII was produced by cooling $H_2O$ at 3 GPa to 77 K and decompressing the sample to ambient pressure.

Ice XII was prepared by compressing $H_2O$ at ≈ 77 K to 1.8 GPa. It thus corresponds to the second regime of metastability for this ice, as defined in Ref. [4].

All samples of polycrystalline ice (5mm diameter spheres) thus obtained were recovered and stored under liquid nitrogen before use. The experiments were carried out using the high energy beamline (Insertion Device 15B) of the European Synchrotron Radiation Facility in Grenoble, France [22]. Simultaneous in-situ x-ray synchrotron diffraction with a MAR image plate set-up was used to check for sample purity and avoid alteration. Rietveld refinements using the FullProf suite [23] were performed to get accurate determination of the lattice parameters of the samples and extract density of each phase. The synchrotron radiation beam was monochromatized to select 86.8 keV photons which were focused on the sample kept inside a temperature cell under vacuum. Cooling was performed using a classical helium displex cryostat which allows maintaining temperature on each measured sample below 10 K during measurements.

Compton scattering spectra were measured using a germanium multidetector with a scattering angle of approximately 160 degrees. Four vertically aligned elements of the detector have been used. In order to avoid parasite scattering coming from the sample environment, a set of carefully designed slits were mounted before and after the sample so as to collimate the incident as well as the scattered radiation, and adjusted in order to confine the scattering volume totally within the sample. The resolution function was deduced from the full width at half maximum (FWHM) of the thermal diffuse scattering peak and is equal to 0.45 a.u. Our statistics range from around $6 \times 10^6$ counts (Ices Ih, VII and XII for around 4 hours of data acquisition each) to $1.4 \times 10^7$ counts (Ices VI and VIII for around 7 hours of data acquisition each) in a bin of 0.03 a.u. at the top of the Compton peak. In this bin, count rate at the top of the Compton peak was around 100 counts per second for each of the pressure point measured. Raw spectra were corrected for background, absorption, and converted into momentum scale. Since we are only interested by the single scattering events, the multiple elastic and inelastic scattering contribution (MSC) was calculated for each measured profile by means of Monte-Carlo simulations taking into account beam polarization, sample geometry and density [24]. MSC was then subtracted from measured profiles. Since the Compton profile is a projection of the electron momentum density, its integral results in the total number of electrons and provides a convenient normalization. Here profiles have been normalized to the number of electrons in a water molecule. Since many of the systematic errors are subtracted out when one takes the difference between two Compton profiles (CPs), the results are presented in form of difference between two profiles, which is thus a very robust quantity.

The program used for the *ab initio* calculations was CRYSTAL98 [25] which is especially suitable for hydrogen-bonded materials [19, 26]. The calculation for ice Ih is described in Ref. [19]. For ice VIII, we have taken the structure from Ref. [27]. The basis set used to express the orbitals is similar to the one used for ice Ih [19]. The occupied orbitals used to determine the electron momentum density and the Compton profiles were calculated within the restricted Hartree-Fock scheme.

## III. RESULTS AND DISCUSSION

Figure 1 presents the difference of CPs measured for ices VI, VII, VIII and XII, using the profile measured for ice Ih as a reference. Two sets of near identical CPs differences can be identified: the first set concerns [ice VIII - ice Ih] and [ice VII - ice Ih] (filled symbols on Fig. 1) and the second set concerns [ice VI - ice Ih] and [ice XII - ice Ih] (opened symbols on Fig. 1). The only theoretical CP difference [ice VIII - ice Ih] matches remarkably with the corresponding experimental one. If the amplitudes of the observed wiggles clearly differ between both sets, the shape is the same with coinciding extrema (at 0, ~ 0.75, ~ 1.35 and ~ 1.8 a.u. in each case). All these ice phases have differing crystalline structure but since we use isotropic polycristalline samples we observe similar oscillatory features representative of local structure, for all phases. In fact, the main contribution to the oscillation is due to an antibonding, repulsive interaction between neighboring water molecules [28, 29, 30]. Other small contributions come from charge polarization [31] and transfer effects [19]. In water the amplitude of the oscillation is also related to the coordination but in all ice polymorphs the water molecules are always hydrogen-bonded to four neighbors within an approximately tetrahedral coordination [2].

Table 1 summarizes the densities of the studied ices as deduced from our Rietveld refinements together with the averaged apparent $r_{OH}$ and $r_{OO}$ lengths, which are the principle structural parameters. The density of the studied ices at similar thermodynamical conditions is the important parameter for the observed amplitudes of the CP differences: the higher is the density with respect to that found in ice Ih, the higher are the amplitudes. Ices VIII and VII have a similar density, as do ices VI and XII, and this leads to the two sets of CP differences identified above. A straightforward explanation stems from the fact that the second moment of the CP is a measure of the electronic kinetic energy [11] and thus a broader profile implies a higher electronic kinetic energy. As ice gets denser, the antibonding, repulsive interaction between neighboring water molecules implies more localized orbitals in position space and an increase in the kinetic energy [28]. This leads to broader CP for ices VI, VII, VIII and XII compared to ice Ih, as seen by the negative CP difference at q = 0 a.u when the Ice Ih CP is subtracted.

Regarding the local structural parameters $r_{OH}$ and $r_{OO}$ (Table 1), both ices VI and XII exhibit very close values, reflected in similar CPs. Ice VII is the fully disordered form of ice VIII, so that both of them have an overall similar structure. Thus, it is not surprising to find similar CPs also for ices VII and VIII. Nevertheless even though $r_{OO}$ is almost the same for both these ices, the apparent $r_{OH}$ in ice VII measured by neutron diffraction is 0.911 [6], i.e. much smaller than in ice VIII. This effect comes from oxygen site disorder. The sites are "displaced by $\delta \approx 0.1$ Å from the average position", leading to "an apparent bond-length shortening of 0.05 Å" [6]. This apparent site disorder is not yet fully understood, and a controversy exists between two possible displacements for the oxygen atom along <111> axes [7] or <100> axes [8]. Our samples being polycrystalline we cannot pinpoint the directional origin of the disorder. However the oxygen site displacement and the apparent $r_{OH}$ are very similar, independent of the directional origin (<111> or <100>). As a consequence, the $r_{OH}$ value given in table 1 is the one given in Ref. [6], but by including the correction of the 0.05 Å defined above, the corrected ice VII $r_{OH}$ value brings the ice VII local parameters close to ice VIII ones and we again conclude that gross similarity in the local structure implies gross similarity in bonding and in the resulting CP.

We now look for finer details in the CPs differences so as to go beyond these initial conclusions. For this we use a new set of CPs differences, using the ordered ice VIII as a reference. Fig. 2 accordingly shows the difference of CPs measured for ice VIII with the CPs of ices VI, VII and XII.

The two CPs differences [ice VIII - ice VI] and [ice VIII - ice XII] are similar, as expected from the earlier considerations of density and local environment dominating the bonding and the CPs. Ice VI and ice XII are of near identical density but are less dense than ice VIII hence their CPs are correspondingly broader leading to a larger negative CP difference at q=0 a.u.

The CP difference between ice VIII and ice VII is worth a closer look. Given the identical densities of the two and similarity of structure, it should be featureless but is clearly non-zero though of small amplitude (less than 0.01% of the maximum of the total profile). The CP difference being grossly determined by the averaged local structure, one can infer that the oxygen disorder in ice VII impacts the electronic structure (in the sense of electronic delocalization) sufficiently so as to induce a CP difference. The oxygen disorder is of course manifest in the position space and fortunately the Fourier transform relating electronic wave functions in position and momentum spaces provides us with a means to confirm this inference. We consider the autocorrelation function B(r), which is the Fourier transform of the CP or the projected momentum density [19]:

$$B(\mathbf{r}) = \int \rho(\mathbf{p}) \exp(-i\mathbf{p}.\mathbf{r}) \, d^3\mathbf{p} \qquad (2)$$

From the convolution theorem, B(r) is just the autocorrelation of the one-electron wave functions. The effect of isotropic oxygen disorder (i.e. 'random' oxygen atom displacement with respect to equlibrium positions by $\delta \approx 0.1$ Å) is simulated by convoluting the B(r) of ice VIII by a Gaussian function with FWHM = 0.1 Å [11] and applying the Fourier transform to simulate oxygen-disordered ice VIII. The inset in Fig. 2 shows the CP difference between ice VIII and a pseudo oxygen-disordered ice VIII (ice VIII$_{Odisorder}$). One finds a wide wiggle with the right phase and sign. At low q the CP difference is negative as in experiment and this finding provides us with a probable explanation of its origin: the oxygen disorder in ice VII.

The effect produced by oxygen disorder can also be illustrated by a comparison between autocorrelation functions of ice VIII, ice VII and calculations for the H$_2$O molecule and ice Ih (Fig. 3). The figure shows differences between autocorrelation functions. An extreme case is the B(r) difference between an isolated water molecule calculation (as representative of ultimate disorder) and that for ice VIII or ice Ih. Hydrogen bonding induces a weakening of the covalent bond and changes in the electronic wave functions due to neighbouring molecules. This is reflected in the calculated differences where the maximum at short length scales implies a stronger covalent bond in the molecule and the minimum at longer length scales is due to perturbations from hydrogen bonding in the condensed state. The experimental differences between ice VII and ice VIII, though understandably much smaller (the phase is condensed matter), show similar behavior with a marked maximum at short length scales and a small minimum at longer length scales. The conclusions are straightforward: oxygen disorder involves a weakening of the hydrogen bonding in ice VII, i.e. a stronger covalent bonding. In other terms, the measured difference between ice VII and ice VIII is due to oxygen site disorder and induces a slightly more molecular nature in the disordered VII phase.

Next, we will quantify the effect of oxygen disorder. Through the CP difference it is possible to measure the fraction of electrons directly involved in a phase transition of this kind, that is the electrons whose wave functions undergo change. The number $n_e$ of electrons involved in this change is defined as follow [17]:

$$n_e = \frac{1}{2}\int_{-\infty}^{+\infty}|\Delta J(q)|\,dq \qquad (3)$$

where $\Delta J(q)$ is the difference between CPs measured for two different forms of ice. In a previous study, $n_e$ allowed us to track specifically the change in the hydrogen bonding coordination in water as a function of the temperature [17]. Table 2 presents the experimental quantities $n_e$ obtained by means of Eq. (3) for the differences between studied ices and ices Ih taken as references. The $n_e$ corresponding to the difference between ice VIII and ice VII is also shown. The quantity $n_e$ relates to change in the local structure of the studied forms of ice and is influenced by at least two different mechanisms, density change and oxygen site disorder. In Table 2. we again find the intuitive result mentioned before where denser phases with respect to the reference phase imply higher amplitudes in the CP difference and a higher value for $n_e$. This trend shows that disregarding differences in structure, the antibonding interaction serves to increase the electronic kinetic energy as density increases. However in considering the CP difference between ices VII and VIII one remarks that this quantity is unexpectedly big given the near identical values found for VII-Ih and VIII-Ih. This is due to the oxygen disorder which induces a change in the CP different from that induced by a density change, as shown in Fig. 2.

In conclusion, we measure changes in Compton profiles for several phases of ice and trace them in good part to increase in electronic kinetic energy with the density of the phase due to the antibonding, repulsive interaction between neighboring water molecules. We also find a good correlation between increase in density and the number of displaced electrons for related forms of ice. The anomalous behavior that we find for ice VII arises from the effect of the oxygen-site disorder.

ACKNOWLEDGMENT

We aknowledge useful discussions of our results with Geneviève Loupias and helpful technical help from Philippe Julien.  Bernardo Barbiellini is supported by the US Department of Energy, Office of Science, Basic Energy Sciences contract DE-FG02 07ER46352 and DE-FG02-08ER46540 (CMSN), and benefited from the allocation of supercomputer time at NERSC and Northeastern University's Advanced Scientific Computation Center (ASCC).

# CAPTIONS.

**FIG. 1.** Experimental Compton profiles (CPs) difference between ice VIII, ice VII, ice VI, ice XII and ice Ih taken as reference. Theoretical CP difference between ice VIII and ice Ih is also shown. In the inset, calculated isolated water molecule and ice Ih CPs are shown, together with their difference. The water molecule CP is narrower than the ice Ih CP showing a higher localization of electronic density in momentum space, i.e. higher delocalization in position space due to the lack of the repulsive antibonding interaction.

**FIG. 2.** Compton profiles differences between ordered ice VIII taken as the reference on the one hand and ice VI, ice XII, ice VII on the other. The inset shows the difference between ice VIII and a pseudo ice VIII taking account of oxygen disordering corresponding to an oxygen atom displacement of 0.1 Å, compared with the difference iceVIII-iceVII.

**FIG. 3.** Autocorrelations functions differences between calculated water molecule and ice VIII (molec-VIII); calculated water molecule and ice Ih (molec-Ih); ices VII and VIII (VII-VIII).

**TABLE I.** Densities and principal structural parameters $r_{OH}$ and $r_{OO}$ for ices Ih, VI, VII, VIII and XII, ranked in order of decreasing densities. Our ices being recovered, densities are given at ambient pressure and are deduced from Rietveld refinements of the diffraction spectra measured for each of our samples. Results are consistent with literature.

**TABLE II.** Experimental fraction of electron $n_e$ obtained for the differences between studied ices and ices Ih taken as reference, together with the $n_e$ corresponding to the difference between ice VIII and ice VII. Error bars are based on statistical uncertainty. We checked that the calculated number of displaced electrons remains the same whatever the upper limit of the integration when chosen above 2.8 a.u.

|  | Ice VIII [6] | Ice VII [6] | Ice VI [32,33] | Ice XII [3,4,34] | Ice Ih |
|---|---|---|---|---|---|
| Density (g.cm$^{-3}$) | 1.50 | 1.50 | 1.32 | 1.29 | 0.92 |
| $r_{OH}$ (Å) - mean value | 0.968 | 0.911 | 0.961 | 0.960 | 0.985 |
| $r_{OO}$ (Å) - mean value | 2.892 | 2.890 | 2.774 | 2.792 | 2.764 |

**TABLE I.**

|  | VIII - Ih | VII - Ih | VI - Ih | XII - Ih | VIII - VII |
|---|---|---|---|---|---|
| $n_e$ (x$10^{-2}$) experiment | **1.44±0.04** | **1.37±0.05** | **0.81±0.05** | **0.77±0.07** | **0.37±0.04** |

**TABLE II.**

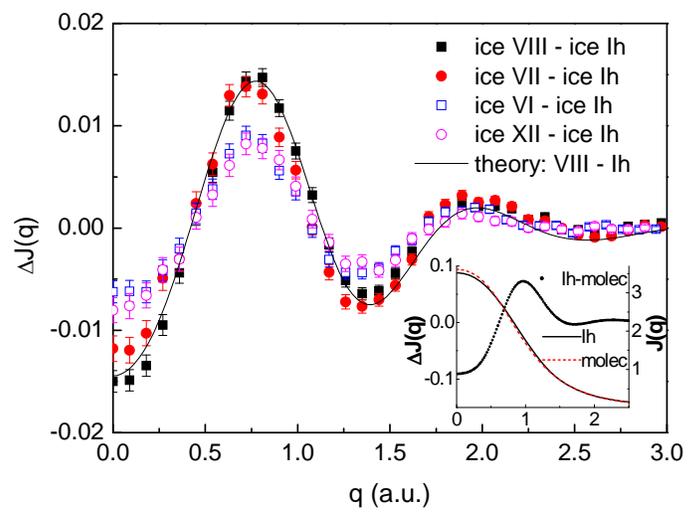

**Fig. 1.**

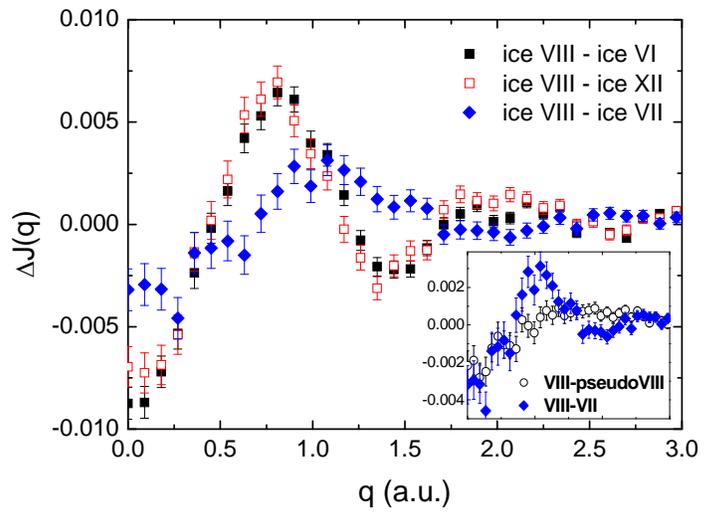

**Fig. 2.**

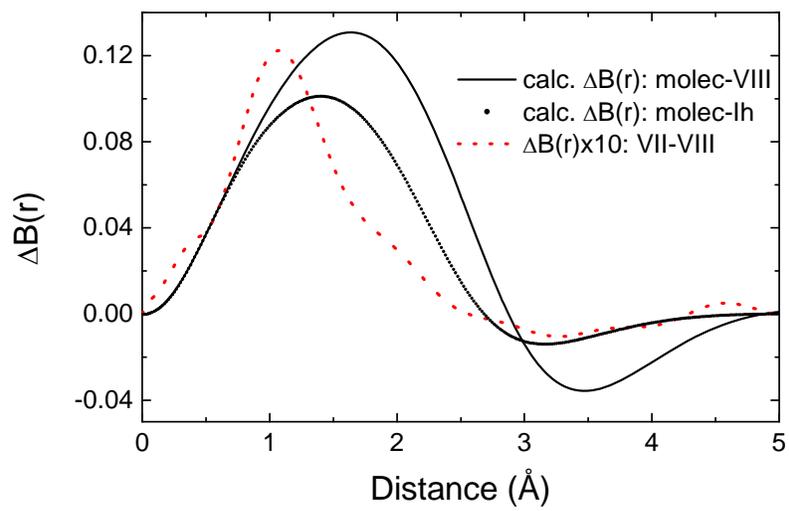

**Fig. 3.**